\documentclass[twocolumn,showpacs,preprintnumbers,amsmath,amssymb]{revtex4}
\usepackage{graphicx}
\usepackage{dcolumn}
\usepackage{bm}
\begin{document}

\title{Self--consistent theory of the long--range order
in solid solutions}

\author{A.~I.~Olemskoi}
\affiliation{Department of Physical Electronics,
Sumy State University, Rimskii--Korsakov St. 2, 40007 Sumy, Ukraine}
\email{olemskoi@ssu.sumy.ua}

\date{\today}

\begin{abstract}
On the basis of the assumption that atoms play a role of effective
Fermions at lattice distribution, the study of the
long--range ordering is shown to be reduced to self--consistent
consideration of single and collective excitations being relevant
to the space distribution of atoms and Fourier transform of such
distribution, respectively. A diagram method advanced allows to elaborate
complete thermodynamic picture of the long--range ordering of the
arbitrary compositional solid solution. The long--range order
parameter is found for different chemical potentials of the
components to obtain a scope of ordering solid solutions according
to relation between degree of the chemical affinity of the
components and mixing energy.
The boundary composition of the ordering phase $AB_n$ is determined
as a function of the chemical potentials of the components and
concentrations of impurities and defects.
Temperature--compositional
dependencies of the order parameter and the sublattice difference
of the chemical potentials are determined explicitly. Polarization
effects and passing out of the compositional domain
$0.318<\overline C<0.682$ is shown make for transformation of the
second order phase transition into the first one. The hydrodynamic
behavior of the system is presented by a reactive mode being
result of the interference of condensate and fluctuation
components of collective excitations. The dispersion law of this
mode is displayed experimentally as the Zener peak of the internal
friction whose frequency and wave number decay
monotonically with temperature increase and phase velocity has a
maximum at intermediate temperatures in ordering domain. The
polarization effects are shown to be relevant to the static
component of Green function, the Goldstone mode of the symmetry
restoration is represented by the instant vertex function.

\end{abstract}

\pacs{05.50.+q, 64.60.Cn, 61.66.Dk}
\keywords{}

\maketitle

\section{Introduction}\label{sec:level1}

In spite of long history, the problem of long--range order in solid
solutions keeps constant interest both of the academic community and
for metallurgy applications \cite{a} --- \cite{D}.
Main peculiarity of contemporary theories along this line consists in using
two marginal approaches being based on consideration either
of the space distribution of atoms of different kinds
or the Fourier transform of such distribution named as concentration waves.
In other words, long--range ordering problem is addressed
to separate considerations of single or
collective excitations distributed over thermodynamic states.
However, it is quite clear that above excitations are strong coupled.
Thus, the problem appears to study systematically both single and collective
excitations in self--consistent manner.
That is main purpose of the present paper that is based on
the methods proposed initially in quantum statistics \cite{9}.

The paper is organized in the following manner.
Section \ref{sec:level2} is devoted to statements
of used formalism whose key point is that
at the lattice sites distribution atoms of a given component
is subjected to a prohibition rule to be considered as effective Fermions.
This allows us to use the well--known formalism
of the second quantization with generic Hamiltonian in a form
inherent in the superconductivity theory.
In Section \ref{sec:level3}, a diagram method is built up
for self--consistent describing both single and collective excitations.
This permits to get general equations for the required set of
the Green functions.
As is shown in Section \ref{sec:level4} devoted to consideration of
the single excitations, their separate description is obtained in
the simplest way within the framework of anomalous quasi--mean value.
First, within this approach, we reproduce shortly main results of standard
mean--field procedure and then we use the Green function
method which allows one to describe not only each of the types of excitations
but take into account their coupling as well. Practical advantage
of this method is a possibility to study in standard manner
solid solutions with arbitrary magnitudes of chemical potential related
to both different components and sublattices of ordering structure.
As a result, we elaborate complete thermodynamic picture of
the long--range ordering of arbitrary solid solution.
Consideration of collective excitations in Section \ref{sec:level5}
allows us to reproduce main details of this picture and to take into account
polarization effects transforming the second order phase transition
into the first one.
The hydrodynamic behavior of the system is shown to be presented by a reactive mode
being result of the interference of condensate and fluctuation components
of collective excitations the last of which is of diffusive type.
Concluding Section \ref{sec:level6} contains a discussion of results
obtained.

\section{Effective Hamiltonian}\label{sec:level2}

We consider a binary solid solution $A$--$B$ with components A and B,
which interact with potential $v_{lm}^{ab}\equiv
v^{ab}({\rm r}_l - {\rm r}_m)$ if components $a, b= A, B$
are placed in lattice sites $l, m$
with coordinates ${\rm r}_l, {\rm r}_m$, correspondingly.
A components distribution over solution sites is given by
microscopic occupation numbers $n_l^A, n_l^B = 0, 1$ being obeyed
to conservation conditions
\begin{eqnarray}
&n_l^A + n_l^B = 1,\quad l = 1, 2,..., N;& \nonumber\\
&\sum_l n_l^a = N_a,\quad a=A, B& \label{1}
\end{eqnarray}
where $N$ is a total number of the solid solution atoms,
$N_A$, $N_B$ are the same for components $A$, $B$
provided $N = N_A + N_B$.
With accounting Eqs. (\ref{1}), the configurational Hamiltonian of the problem
\begin{equation}
{\cal H} = {1 \over 2}\sum_{lm}\sum_{ab} v_{lm}^{ab} n_l^a n_m^b
- \sum_{al}\mu_a n_l^a
\label{2}
\end{equation}
where $\mu_a$ is chemical potential of the components $a$,
is reduced to usual form
\begin{eqnarray}
&{\cal H} = {\cal H}_0 + {\cal H}_1 + {\cal H}_2;&
\label{3}\\
&{\cal H}_0 \equiv {1 \over 2}\sum_{lm}
v_{lm}^{BB} - \mu_B N,&
\label{4}\\
&{\cal H}_1 \equiv \sum_{l}(\varepsilon_l - \mu) n_l,\quad
\varepsilon_l\equiv\sum_{m}
(v_{lm}^{AB} - v_{lm}^{BB}),&
\nonumber\\
&\mu\equiv \mu_A - \mu_B,\quad n_l\equiv n_l^A,&
\label{5}\\
&{\cal H}_2 \equiv {1 \over 2}\sum_{lm} w_{lm} n_l n_m,
\ w_{lm}\equiv v^{AA}_{lm} + v^{BB}_{lm} - 2v^{AB}_{lm}.\ &
\label{6}
\end{eqnarray}
Hereafter, the only $A$--component occupation number
$n_l\equiv n_l^A$ will be used.

The corner stone of our approach is that, in sense of the lattice
sites distribution, atoms of a given component is subjected to a
prohibition rule $n_l = 0$ or $n_l = 1$ to be considered as
effective Fermions distributed over "states" $l$ \cite{KO}. Thus,
it is convenient to use the well--known formalism of the second
quantization. With this aim, we represent the occupation numbers
in the usual form $n_l\equiv a^{+}_l a_l$ where creation and
annihilation operators $a^{+}_l, a_l$ are subjected to the
anticommutation rules $\{a_l, a^{+}_m\} = \delta_{lm}$ (as usual,
braces denote anticommutator). Moreover, the ordering process
causes lattice splitting into two sublattices which will be
labelled by Greek indexes 0 or 1. As a result,
Hamiltonian $H\equiv {\cal H} - {\cal H}_0$ counted out the
non--essential constant ${\cal H}_0$ takes the standard second
quantization form:
\begin{equation}
H = {1\over 2}\sum_{l\alpha}|\varepsilon_l-\mu_\alpha| a^{+}_{l\alpha}a_{l\alpha} +
{1\over 2}\sum_{lm} w_{lm} a^{+}_{m 1} a^{+}_{l 0}a_{m 0} a_{l 1}.
\label{7}
\end{equation}
Here, the anticommutation properties and natural condition
$w_{ll}\equiv 0$ (a self--action is absent) take into account,
the expressions for a Fermion effective energy $\varepsilon_l$,
its chemical potential $\mu_\alpha$
(being dependent on sublattice number $\alpha$)
and an effective interaction potential $w_{lm}$
are defined in Eqs. (\ref{5}), (\ref{6}).
The key point is that atoms $A$
on their own sublattice $\alpha=0$ and on outsider one $\alpha=1$,
which chemical potentials are subjected to
the condition $\mu_0<\varepsilon_l<\mu_1$, are considered
in similar manner as particle and hole in semiconductors.
On the other hand, we are kept in last term the only contribution
of the sublattices that causes the ordering process \cite{KO}.

Sometimes, it is convenient to use instead of the above site representation
the wave one:
\begin{equation}
a_{\rm k} \equiv \sqrt{2\over N}\sum_{l} a_l e^{-{\rm i}{\bf k}{\bf r}_{l}},
\ \
a^{+}_{\rm k} \equiv \sqrt{2\over N}\sum_{l} a^{+}_l e^{{\rm i}
{\bf k}{\bf r}_{l}},
\label{8}
\end{equation}
\begin{equation}
\varepsilon_{\bf k}\equiv {2 \over N}\sum_{l} \varepsilon_{l}
e^{{\rm i}{\bf k}{\bf r}_{l}},\quad
w_{\bf k}\equiv {2\over N}\sum_{lm} w_{lm}
e^{{\rm i}{\bf k}\left({\bf r}_{l} - {\bf r}_{m}\right)}
\label{9}
\end{equation}
where summation is fulfilled over $N/2$ sites of a sublattice.
Then, the expression (\ref{7}) takes the form
\begin{widetext}
\begin{equation}
H = {1\over 2}\sum_{{\bf k}\alpha}|\varepsilon-\mu_\alpha|
a^{+}_{{\bf k}\alpha}a_{{\bf k}\alpha} +
{1 \over N}\sum_{{\bf k}{\bf k'}{\bf q}} w_{{\bf k}-{\bf k'}}
a^{+}_{{\bf k} 1} a^{+}_{{\bf -k}+
{\bf q}, 0} a_{{\bf k'} 0} a_{{\bf -k'}+{\bf q}, 1}.
\label{10}
\end{equation}
Here, we take into account the lattice translational invariance condition
according to which the definition (\ref{5}) derives to a constant
Fermion energy $\varepsilon_l\equiv\varepsilon$ so that
$\varepsilon_{\bf k}=\varepsilon\delta_{{\bf k}0}$.
\end{widetext}

\section{Development of the self--consistent scheme}\label{sec:level3}

For describing both single and collective excitations, it is convenient
to proceed from different (site or wave) representations, so that we
shall first make recourse to the diagram method making no use of the
explicit form  of the appropriate Hamiltonian. This will permit to get the
general view of equations for the required set of the Green functions.

Description of the present
system is ensured by the use of Matsubara's Green function
\begin{equation}
G_{lm}^{\alpha \beta}(t)= -\left< \widehat{T}a_{l\alpha}(t)
a_{m\beta}^+(0) \right>
\label{10a}
\end{equation}
where $a_{l\alpha}^+(t)$, $a_{l\alpha}(t)$
are the creation and annihilation operators within the Heisenberg
representation, $t$ is the imaginary time,
the angular brackets mean averaging over configurational states
of the solid solution (see \cite{KO}), the other
symbols are standard  \cite{9}. Making use of the above anticommutation
relations, it is easy to convince that the normal
(diagonal) and anomalous (off--diagonal) components possess the
following properties:
\begin{eqnarray}
-G_{ml}^{11}(-t)=G_{lm}^{00}(t)&\equiv &G_{lm}(t),\nonumber \\
G_{ml}^{10*}(t)=G_{lm}^{01}(t)&\equiv &F_{lm}(t).
\label{11}
\end{eqnarray}
Thanks to this, it is sufficient to examine the behavior of only
two components --- the normal $G_{lm}(t)$ and the anomalous $F_{lm}(t)$
being shown in Figures 1a, b, respectively. Further, when making general
computations, it is also convenient to use the matrix
representation of the type $\widehat{G}_{lm}(t)$ where the cap
signifies the exhaustive search for the $\alpha$ and $\beta$
indices.
\begin{figure}[htb]
\caption{Diagram representation of single and collective
excitations}
\end{figure}

In addition to the matrix nature of the Green function itself, the
presence of the sublattices $\alpha, \beta =0$,~1  results in
matrix structure of the four--tail vertices of the effective
interaction depicted in Figure 1c (here, solid line corresponds
to a solution component on own sublattice ($\alpha, \beta = 0$)
and on another's one ($\alpha, \beta = 1$), the dashed line is
respective for the mixture potential $w_{lm}$). Since the ordering
is caused only by the coupling between components belonging to
different sublattices, for fixed indices $l,\ m$ we shall consider
as non--zero only the components $w_{01,01}\equiv w_{00}$, \
$w_{10,10}\equiv w_{11}$, \ $w_{01,10}\equiv w_{01}$, \
$w_{10,01}\equiv w_{10}$ producing the $\widehat{w}$  matrix of
the second rank (here, one means that interaction
$w_{\gamma\delta,\alpha\beta}$ transforms the sublattices
$\alpha\beta$ into $\gamma\delta$). The $\widehat{\Gamma}$ matrix
(see Figure 1d) of the vertex function is structured in the similar
manner. However, while the $\widehat{w}$ matrix of the bare
interaction is evidently diagonal, the complete vertex
$\widehat{\Gamma}$ has, as will be seen below, all
nonzero--components.

In the diagram representation the Dyson matrix equation takes an
ordinary form depicted in Figure 1e where the thick line
corresponds to the exact Green function, and the thin line
does to the bare one (the matrix of the latter is diagonal).
The self--energy function (Figure  1f) is expressed by the
equation illustrated in Figure 1g. As a result, the problem reduces to the
self--consistent determination of the vertex function
$\widehat{\Gamma}$. It can be shown in standard manner \cite{9} that, at
the expansion in terms of $\widehat{w}$, the main contribution is
made by the terms containing the polarizer $\widehat{\Pi}$ (see Figure
1h) the matrix components $\Pi_{\alpha \beta}$ of which are
determined in a similar way to the $w_{\alpha \beta}$ and $\Gamma
_{\alpha \beta}$. Then the appropriate series reduces to the ladder
sequence which can be represented in the form of the Bethe--Salpeter
equation shown in Figure 1i. It closes the system of equations for the
self--consistent description of the ordering process in solid solution.

In the analytic representation, this system is written as follows:
\begin{eqnarray} &&\widehat{G}^{-1}(\omega
_s)=\left[i\omega_s\widehat{\delta} + (\widehat{\mu}-
\widehat{\varepsilon})\right]
-\widehat{\Sigma}(\omega _s),\label{12a}\\
&&\Sigma_{\alpha\beta}(t)=G_{\beta\alpha}(-t)
\Gamma_{\alpha\beta}(t),\label{12b}\\
&&\widehat{\Gamma}^{-1}(\Omega_S )=\widehat{w}^{-1} -
\widehat{\Pi}(\Omega_S),\label{12c}\\ &&\Pi_{\alpha\beta}(t) =\left[
G_{\alpha\beta}(t) \right]^2.\label{12d} \end{eqnarray}
\noindent
Here, the frequencies $\omega_s=\pi (2s+1)T$, $\Omega _S=2\pi ST$
of the single and collective excitations, respectively, are inherent in
the Fermi and Bose particles, $T$ is temperature in energy units,
$s,\> S=0,\> \pm 1,\>\ldots$ are integers;
$\widehat{\delta}$ is a unit matrix,
$\widehat{\varepsilon}$ is a diagonal matrix with the
elements $\varepsilon^{00}= -\varepsilon^{11}=\varepsilon$,~
$\widehat{\mu}$ is the same with the elements
$\mu^{00}=\mu_0$, $\mu^{11}=-\mu_1$
($\varepsilon$, $\mu_\alpha$, $\alpha=1, 2$ are the bare energy and
the sublattice chemical potentials of a Fermion).
The matrix structure of the
Green function (\ref{12a}) is stated on the Fermi conditions (\ref{11}).
With derivation of the equation (\ref{12c}) for the vertex function
$\widehat{\Gamma}$, it is generally taken
that the bare potential reduces to the constant $\widehat{w}$.
A distinctive feature of the system obtained consists in the fact that
the explicit expressions (\ref{12a}), (\ref{12c}) for the Green
functions $\widehat{G}$, $\widehat{\Gamma}$ of the single and collective
excitations, respectively, are obtained in the frequency representation,
while the expressions (\ref{12b}) and (\ref{12d}) for the
self--energy function $\widehat{\Sigma}$ and polarizer
$\widehat{\Pi}$ require the application of the time representation.
As far as the site and wave representations are concerned, their
choice depends on the type of excitations.

\section{Single excitations}\label{sec:level4}

If we are interested only in the behavior of the single
excitations, their description is obtained in the simplest way
within the framework of the quasi--mean value method \cite{35}.
Applying the standard procedure, it can be shown on the basis of
the expression (\ref{7}) that in the limit $N \to \infty$ the
behavior of the system is asymptotically defined by the
approximating Hamiltonian which takes, in the self--consistent
field approximation, the following form:
\begin{equation}
{\cal H}_{ef}\equiv{1\over 2}\sum_{l\alpha}|\varepsilon_l-\mu_\alpha|
a^+_{l\alpha} a_{l\alpha}
-{1\over 2}\sum_l \left(\Delta_l^* a^+_{l 0}a_{l 1}+\Delta_l a^+_{l 1}a_{l 0}\right).
\label{13}
\end{equation}
Anomalous quasi--mean values
\begin{equation}
\eta_l\equiv\left< a^+_{l 0}a_{l 1}\right>,\quad
\eta^*_l\equiv\left< a^+_{l 1}a_{l 0}\right>
\label{14}
\end{equation}
determining the amplitude of the effective Fermion
transfer from one sublattice to another represent
the (local) order parameter which value gives a gap
\begin{equation}
\Delta_l\equiv-{1\over2}\sum_m w_{lm} \eta_m,\quad
\Delta_l^* \equiv-{1\over2}\sum_m w_{lm} \eta_m^*.
\label{15}
\end{equation}
Here, multipliers $1/2$ in front of sums appear to take summation
over whole lattice sites (but not over sublattice ones). In case
of the long--range ordering, the order parameter,
given Eq. (\ref{14}), becomes constant $(\eta_l\to\eta)$ and, within the usual
mean--field approximation, Eqs. (\ref{15}) give the simple
expression for the gap:
\begin{equation}
\Delta ={W\over 2}\eta,\quad W\equiv-\sum_m w_{lm}\approx -zw
\label{16}
\end{equation}
where the last equality is taken within approximation of
interaction of near neighbors which mixing potential is $w$ and
number is $z$. For case of the short--range ordering, the
correlation parameter (\ref{14}) slows down with $l$ very fast.

Diagonalization of Hamiltonian (\ref{13}) is achieved by means of the
transformations
\begin{eqnarray}
\alpha_{l+}=u_l a^+_{l 1} + v_l a^+_{l 0}\>,\qquad
\alpha_{l-}=u_l a_{l 0} - v_l a_{l 1};\nonumber\\
a_{l 0} = u_l \alpha_{l-} + v_l \alpha_{l+}^+,\qquad
a_{l 1} = u_l \alpha_{l+}^+ - v_l \alpha_{l-}
\label{17}
\end{eqnarray}
where, by virtue of the anticommutation rule,  $u^2_l+v^2_l=1$.
It results in giving
\begin{widetext}
\begin{eqnarray}
{\cal H}_{ef}=U+{1\over 2}\sum_l \epsilon_l
\left( \alpha^+_{l-}\alpha_{l-}+ \alpha^+_{l+}\alpha_{l+}\right)
+{1\over 2}\sum_l \widetilde\mu\left(
\alpha^+_{l-}\alpha_{l-}-\alpha^+_{l+}\alpha_{l+}\right)
\label{18}
\end{eqnarray}
\end{widetext}
where we are restricted ourselves with the case of a real order
parameter $(\eta_l=\eta_l^*)$ and one denotes:
\begin{eqnarray}
&U= {N\over 2}\widetilde\mu-{1\over 2}\sum_l\epsilon_l,\quad
\widetilde\mu\equiv{1\over 2}(\mu_1-\mu_0);& \nonumber \\
&\epsilon_l=\sqrt{(\bar\mu-\varepsilon)^2+\Delta^2_l}\>,\quad
\bar\mu\equiv{1\over 2}(\mu_0+\mu_1);&
\label{19}\\
&\left. {u^2_l\atop
v^2_l}\right\}={1\over 2}\left( 1\mp {\bar\mu-\varepsilon\over
\epsilon_l}\right).&\nonumber
\end{eqnarray}
The quantity $U$ represents the energy of the ground state
whose wave function is
\begin{equation}
\left.  |\Psi_0\right> =\prod_l \left(
u_l+v_l a^+_{l 1}a_{l 0}\right)\left.  |0\right>
\label{20}
\end{equation}
where $\left. |0\right>$ is the wave function of the Fermi vacuum.
In general case of arbitrary concentration $\overline C$ of solid
solution, excitations defined by the  operators  $\alpha_{l+}$,
$\alpha_{l-}$ are non--coincident. Postponing this case below, let
us consider at first the simplest case of stoichiometric solution
($\overline C = 1/2$) where the coincidence of the behavior of
elementary excitations in the both states corresponding to
sublattices $\alpha=0, 1$ is observed ($\alpha_{l+}\equiv
\alpha_{l-}\equiv\alpha$). It is easy to see that in this case
$\widetilde\mu=0$.

The equation for the gap is obtained by substitution of the
equalities (\ref{17}) in the definitions (\ref{14}), (\ref{15}):
\begin{equation}
-{1\over2}\sum_m w_{lm}~{\Delta_m\over \epsilon_m}~
\tanh {\epsilon_m\over 2T} = \Delta_l,\
\epsilon_m\equiv\sqrt{(\bar\mu-\varepsilon)^2+\Delta^2_m}.
\label{21}
\end{equation}
Here, we have used Fermi distribution for excitations number
\begin{equation}
\nu_l\equiv \left<\alpha^+_l\alpha_l\right>
= \left[1+\exp(\epsilon_l/T)\right]^{-1}.
\label{22}
\end{equation}
In the case of the long--range ordering, when dependence of the gap
$\Delta_l\to\Delta$ on the site number $l$ disappears,
we obtain the usual equation within the mean--field approximation:
\begin{equation}
\tanh {\epsilon\over 2T} ={2\epsilon\over W},\
\epsilon\equiv\sqrt{(\bar\mu-\varepsilon)^2+\Delta^2},\
W\equiv -\sum_m w_{lm}.
\label{23}
\end{equation}
Numerical solution of integral equation (\ref{21}) allows us to
take into account effects of short--range correlations as well, for which
the parameter $\Delta_l$ is a strong function of the site number $l$.

Concentrations $C_{l \alpha}\equiv
\left<a^+_{l \alpha}a_{l \alpha}\right>$
of $A$--type component on the sublattices
$\alpha=0, 1$ is determined by Eqs. (\ref{17}), (\ref{19}), (\ref{22}):
\begin{eqnarray}
\left. C_{l 0}\atop C_{l 1}\right\}=
{1\over 2}\left(1\mp{\bar\mu-\varepsilon\over\epsilon_l}
\tanh {\epsilon_l\over 2T}\right).
\label{24}
\end{eqnarray}
On the other hand, the difference $\widetilde\mu$
of the chemical potentials on the sublattices
is given by the condition \cite{37}
\begin{widetext}
\begin{eqnarray}
&\partial\Omega/\partial\widetilde\mu\equiv - N \overline C;&\nonumber\\
&\Omega\equiv -T \ln\left<\exp\left(-\frac{{\cal H}_{ef}-N\tilde\mu}{T}
\right)\right>,
\quad \overline C\equiv {1\over 2N}\sum\limits_l(C_{l 0}+C_{l 1}).&
\label{25}
\end{eqnarray}
\end{widetext}
Making use of Eqs. (\ref{18}) gives
\begin{equation}
\Omega = (U-N\tilde\mu)+T\sum\limits_l\ln (1-\nu_l)
\label{26}
\end{equation}
and thermodynamic equalities (\ref{25}) arrive at trivial result
$\overline C = 1/2$ for mean concentration.
What about the entropy $S\equiv-\partial\Omega/\partial T$, thermodynamic
potential (\ref{26}) derives to usual configuration form
\begin{equation}
S=-\sum\limits_l\left[\nu_l\ln\nu_l+(1-\nu_l)\ln(1-\nu_l)\right].
\label{26a}
\end{equation}

With transition to the ordered state, the jump of
the thermodynamic potential $\Delta F\equiv F_{ord}-F_{disord}$ is
determined by the formula
\begin{equation}
\Delta F=-{1\over2W}\sum_l\Delta_l^2
\label{27}
\end{equation}
following from the equalities  $\partial F/\partial\Delta_l
=\left< \partial {\cal H}_{ef}/\partial \Delta_l\right> $
and (\ref{18}), (\ref{19}), (\ref{22}), (\ref{23}).
According to (\ref{27}) during ordering, as it usually is during
transition to the low--symmetry phase, value $F$ decreases.

The elementary excitation energy  ${\cal E}_{{\rm ex}}\equiv\sum_l
\epsilon_l\nu_l$  is at $\overline C=1/2$ of the form:
\begin{equation}
{\cal E}_{{\rm ex}}=
\sum_l{\sqrt{(\bar\mu-\varepsilon)^2+\Delta_l^2}\over
1+\exp\left[\sqrt{(\bar\mu-\varepsilon)^2+\Delta_l^2}/T\right]}.
\label{28}
\end{equation}

Previously, we applied the mean--field method bringing in
shortest ways to the description of single excitations. Naturally,
the illustrated results can also be obtained by means of the Green function
method which allows for determining the Fermi function $\widehat{G}(\omega_s)$  from the
system (\ref{12a}) --- (\ref{12d}).
Keeping in mind the fact that the Green procedure
allows for describing not only each of the types of excitations but for
taking into account their coupling, we shall demonstrate at first the way it
reproduces the results of the mean--field method.

In the context of the above used site representation,
the functions $\widehat{G}$, $\varepsilon$
and $\widehat{\Sigma}$ in
the equation (\ref{12a}) and functions $\widehat{\Gamma}$,
$\widehat{w}$ and $\widehat{\Pi}$ in the equation (\ref{12c}) should
be corresponded the lattice index $l$, and factorization of the appropriate
contribution (\ref{12b}) and (\ref{12d}) is achieved in transition to the wave
representation.

To define the explicit form of the Green function (\ref{12a}), we shall
consider that, as in the case of the theory of superconductivity, the
self--energy function $\widehat{\Sigma}_l$  acquires,
under ordering, the off--diagonal
components  $\Sigma_{01~l} =\Sigma^*_{10~l} =\Delta_l$
corresponding to the gap  $\Delta_l$  in the energy
spectrum of the single excitations \cite{9}.
The matrix $\widehat{G}_l$ inversion
leads in the case $\overline C=1/2$ to the following expressions:
\begin{widetext}
\begin{eqnarray}
G_l(\omega_s)=-{{\rm i}\omega_s+(\bar\mu-\varepsilon) \over
\omega_s^2+\epsilon_l^2}\>, \ \ F_l(\omega_s)=-{\Delta_l \over
\omega_s^2+\epsilon_l^2}\>;\quad
\epsilon_l=\sqrt{(\varepsilon-\bar\mu)^2+\Delta_l^2},\ \ \omega_s=\pi
(1+2s)T, \ s=0, \pm 1, \ldots
\label{29}
\end{eqnarray}
\end{widetext}
With accounting the sum rules \cite{43}
\begin{equation}
T\sum_{s=-\infty}^{\infty}
{1\over\omega_s^2+\epsilon^2}=
{1\over 2\epsilon}\tanh {\epsilon\over 2T},\
\sum_{s=-\infty}^{\infty}
{\omega_s\over\omega_s^2+\epsilon^2}=0
\label{36}
\end{equation}
being applicable for some real $\epsilon$, substitution of Eqs. (\ref{29})
into Eq. (\ref{12b}) for the off--diagonal
components, where the vertex $\widehat{\Gamma}_l$ is
reduced to convolution operator of the bare potential
${\widehat w}_l\equiv\sum_m w_{lm}$,
arrives at the self--consistency equation which, as would be expected,
coincides with Eq. (\ref{21}) for the gap $\Delta_l$.
It is key point that the diagrammatic approach allows to approve
the self--consistency solution of Eq. (\ref{21}) by means of using some
approximation for the vertex $\widehat{\Gamma}_l$ (see Section
\ref{sec:level5}).
Dependence of the effective chemical potential  $\widetilde\mu$
on the difference of the lattice concentrations  $C_{l \alpha}$
follows from the relation
\begin{equation}
\widetilde C_l\equiv{1\over 2}(C_{l1}-C_{l0})=-T\sum_s G_l(\omega_s)
\label{30aa}
\end{equation}
which, with the application of
the first one from the equalities (\ref{29}), leads to result
\begin{equation}
\widetilde C_l=\frac{\bar\mu-\varepsilon}{2\epsilon_l}
\tanh {\epsilon_l\over 2T}
\label{30b}
\end{equation}
following from Eq. (\ref{24}) as well.

Polarizer (\ref{12d}) has the following  Fourier--form:
\begin{equation}
\Pi^{\alpha\beta}_l(\Omega_S)=-T\sum_{s=-\infty}^\infty {1\over N}
\sum_m G^{\alpha\beta}_{l-m}(\Omega_S-\omega_s)
G^{\alpha\beta}_m(\omega_s).
\label{30}
\end{equation}
Substituting the Green functions (\ref{29}) here, upon summing over $s$, we
arrive at the expressions
\begin{widetext}
\begin{eqnarray}
&\Pi^{00}(0)=\Pi^{11}(0)=-{1\over 2N}\sum\limits_l\epsilon_l^{-1}{\rm
tanh}{\epsilon_l\over 2T}+A\Delta^2,\quad
\Pi^{01}(0)=\Pi^{10}(0)=-A\Delta^2;&
\label{31}\\
&A\equiv {1\over 4N}\sum\limits_l\epsilon_l^{-3}
\left(\tanh {\epsilon_l\over 2T}
-{\epsilon_l\over 2T}{\rm cosh}^{-2}{\epsilon_l\over 2T}\right),&
\nonumber
\end{eqnarray}
\end{widetext}
where, in view of the macroscopic equivalence of sites, there is
no dependence on their number $l$. Making use of Eq. (\ref{30}), we
find the inverse vertex function (\ref{12c}) responsible
for the behavior of the collective excitations (see Section \ref{sec:level5}).

The system of equations (\ref{14}), (\ref{15}), (\ref{21}),
(\ref{25}) and (\ref{26}) that offers the self--consistent
description of the single excitations affords finding the gap
$\Delta$, the order parameter $\eta$ and the sublattice difference
of concentrations $\widetilde C$ by the assigned values of
thermodynamic parameters (the temperature $T$ and the chemical
potential $\bar\mu$), as well as microscopic parameters (the value
of the mixing potential $w_{lm}$ and effective energy
$\varepsilon$ given by Eqs. (\ref{5}), (\ref{6})). Numerical
solution of these equations for the long--range order parameter
results in the temperature dependencies displayed in Figure 2.
\begin{figure}[htb]
\caption{Temperature dependencies of the long--range order parameter
at different magnitudes of chemical potentials (curves top-down correspond
to values $(\bar\mu-\varepsilon)/\Delta_0=0.0,\ 0.2,\ 0.3,\ 0.4,\ 0.5,\ 0.6,\ 0.7,\
0.8,\ 0.9$. Concentration $\overline C=0.5$)}
\end{figure}
From this figure we notice that the parameter $\eta$
monotonically decays as the temperature $T$ and
the difference $|\bar\mu-\varepsilon|$ increase.
According to Eqs. (\ref{23}), (\ref{30b}),
at long--range ordering the sublattice concentration difference is
\begin{equation}
\widetilde C = {\bar\mu-\varepsilon\over W}
\label{30a}
\end{equation}
to be determined by the mean chemical potential but being
non--dependent on temperature. Usually, one supposes
$\bar\mu=\varepsilon$ and, as consequence, $\widetilde C=0$ (i.e.,
$C_0=C_1$).

The ground system state defined by the function (\ref{20}) is
achieved at $T = 0$ where site dependence is missing and
excitations are absent ($\nu=0$). At $\bar\mu=\varepsilon$ the gap
width and order parameter take maximum values $\Delta_0={1\over
2}W\equiv-{1\over 2}\sum_m w_{lm}$, $\eta=1$, variation of the
ground state energy, being equal to the decreasing thermodynamic
potential, takes maximum value $|\Delta F|={N\over 4}\Delta_0$;
respectively, the sublattice concentrations $C_{0}$, $C_{1}$
coincide. With finite difference $\bar\mu-\varepsilon\ne 0$, there
are $\Delta=\sqrt{\Delta_0^2-(\bar\mu-\varepsilon)^2}$,
$\eta=\sqrt{1-(\bar\mu-\varepsilon)^2/\Delta_0^2}$, $\Delta
F=-{N\over
4}\Delta_0\left[1-(\bar\mu-\varepsilon)^2/\Delta_0^2\right]$,
$\widetilde C=(\bar\mu-\varepsilon)/2\Delta_0$. Thus, as the
sublattice chemical potentials $\mu_0=\mu_1=\bar\mu$ increases
from value $\varepsilon$ to $\varepsilon+\Delta_0$, the gap width,
order parameter and absolute value of the system free energy
change monotonically decay going into zero at the boundary value
$\bar\mu_c=\varepsilon+\Delta_0$. The sublattice concentrations
difference varies here linearly from 0 to $1/2$.

As the temperature arises, originated are the elementary
excitations the number of which $\nu\approx\exp (-\Delta_0/T)\ll 1$.
With a precision of
the first power terms over $\nu\ll 1$, we obtain:
\begin{widetext}
\begin{eqnarray}
&\Delta=\Delta_0\sqrt{1-{\left(\bar\mu-\varepsilon\over\Delta_0\right)^2}}
\left\{1-2\left[1-{\left(\bar\mu-\varepsilon\over\Delta_0\right)^2}\right]^{-1}
\exp\left(-{\Delta_0\over T}\right)\right\},&\nonumber\\
&\eta={\Delta\over \Delta_0},
\quad \widetilde C={\bar\mu-\varepsilon\over 2\Delta_0},\quad
\Delta_0={1\over 2}W\equiv-{1\over 2}\sum\limits_m w_{lm},&
\label{32}\\
&\Delta F=-{N\over 4}\Delta_0\eta^2,\quad
{\cal E}_{\rm ex}=N\Delta_0\exp\left(-{\Delta_0\over T}\right).&
\nonumber
\end{eqnarray}
As it is known, the exponential character is inherent in the
low--temperature dependencies usually observed under phase
transitions.

Near critical temperature $T_c=W/4$ the ordering process can be
presented in analytical form within the mean--field approximation
that is stated on making use of Eq. (\ref{23}). Within domain
$0<(T_c-T)/T_c\ll 1$, we obtain:
\begin{eqnarray}
&\Delta=2T_c\sqrt{3t},\quad \eta=\sqrt{3t},\quad
t\equiv{T_c-T\over T_c},\quad
T_c ={1\over 4}W,\quad W\equiv-\sum_m w_{lm},&\nonumber\\
&\Delta F=-{3\over 2}NT_c t,\quad
{\cal E}_{\rm ex}=NT_c\sqrt{3t}&
\label{33}
\end{eqnarray}
where we put $\bar\mu=\varepsilon$.

Now, we consider general case of non--stoichiometric solid
solution which the (mean) concentration $\overline C\ne 1/2$ and
the difference $\widetilde\mu$ of the sublattice chemical
potentials is determined by equations (cf. Eq. (\ref{25}))
\begin{eqnarray}
&\partial \Omega/\partial \widetilde\mu\equiv -N \overline C;&
\nonumber\\
&\Omega=-{N\over 2}\widetilde\mu-{1\over 2}\sum\limits_l\epsilon_l
+T\sum\limits_l\ln (1-\nu_{l+})+T\sum\limits_l\ln (1-\nu_{l-}),&
\label{34} \\
&\nu_{l\pm}=\left\{1+\exp\left(\epsilon_l\mp\widetilde\mu\over 2T\right)
\right\}^{-1}&
\nonumber
\end{eqnarray}
following from Eqs. (\ref{25}), (\ref{18}) and (\ref{19}).
Respectively, inversion of matrix (\ref{12a}) gives instead of Eqs. (\ref{29})
\begin{eqnarray}
&G^{00}_l(\omega_s)=\left[{\rm i}\omega_s-(\mu_1-\varepsilon)\right]/
{\cal D}_l, \quad
G^{11}_l(\omega_s)=\left[{\rm i}\omega_s+(\mu_0-\varepsilon)\right]/
{\cal D}_l,&\nonumber\\
&G^{01}_l(\omega_s)= \Delta_l/{\cal D}_l,\ \
G^{10}_l(\omega_s)=\Delta_l^*/{\cal D}_l;\quad
{\cal D}_l(\omega_s)\equiv
\left[{\rm i}\omega_s-(\epsilon_l+\widetilde\mu)\right]
\left[{\rm i}\omega_s+(\epsilon_l-\widetilde\mu)\right],&\label{35}\\
&\epsilon_l=\sqrt{
(\bar\mu-\varepsilon)^2 +|\Delta_l|^2},\quad
\bar\mu\equiv{1\over 2}(\mu_0+\mu_1),~~\widetilde\mu\equiv
{1\over 2}(\mu_1-\mu_0),&
\nonumber\\
&\omega_s=\pi (1+2s)T, \quad s=0, \pm 1, \ldots &\nonumber
\end{eqnarray}
With accounting Eqs. (\ref{36}) and Eq. (\ref{12b})
with $\widehat\Gamma$ substituted by $\widehat w$, we obtain
the self--consistency condition
\begin{equation}
-{1\over4}\sum_m w_{lm}~{\Delta_m\over \epsilon_m}~
\left(\tanh {\epsilon_m+\widetilde\mu \over 2T}+
\tanh {\epsilon_m-\widetilde\mu \over 2T}\right)=\Delta_l
\label{37}
\end{equation}
replacing Eq. (\ref{21}).
Respectively, from equalities $C_{\alpha}\equiv
-T\sum_s G_{\alpha\alpha}(\omega_s)$ one follows:
\begin{eqnarray}
C_{0} =
{\epsilon-(\bar\mu-\varepsilon)\over 4\epsilon}
\tanh {\epsilon+\widetilde\mu \over 2T}-
{\epsilon+(\bar\mu-\varepsilon)\over 4\epsilon}
\tanh {\epsilon-\widetilde\mu \over 2T},
\nonumber\\
\label{38}\\
C_{1} =
{\epsilon+(\bar\mu-\varepsilon)\over 4\epsilon}
\tanh {\epsilon+\widetilde\mu \over 2T}-
{\epsilon-(\bar\mu-\varepsilon)\over 4\epsilon}
\tanh {\epsilon-\widetilde\mu \over 2T}.
\nonumber
\end{eqnarray}
In case of the long--range ordering, when the parameters
$\Delta_l\to\Delta$, $\epsilon_l\to\epsilon$ do not depend on the site
number $l$, the relations (\ref{37}), (\ref{38})
arrive at Eq. (\ref{30a}), as follows.
On the other hand, thermodynamic equality (\ref{34}) derive to equation
\begin{eqnarray}
2\overline C - 1={1\over N}\sum\limits_l\left(\nu_{l+}-\nu_{l-}\right)
\equiv{1\over 2N}\sum\limits_l\left(
\tanh {\epsilon_l+\widetilde\mu\over 2T}-
\tanh {\epsilon_l-\widetilde\mu\over 2T}\right)
\label{39}
\end{eqnarray}
where the relation
$1-2\nu_{l\pm}=\tanh {\epsilon_l\mp\widetilde\mu\over 2T}$
is taken into account.
In difference of Eq. (\ref{25}), this equation is non--trivial
and accompanied with Eqs. (\ref{37}), (\ref{38}) determines
quantities $\Delta_l$, $C_0$, $C_1$, $\widetilde\mu$
at given thermodynamic parameters $T$, $\overline C$, $\bar\mu$
and microscopic ones $w_{lm}$ and $\varepsilon$.
In similar manner, the excitation energy is determined by
the expressions (cf. Eq. (\ref{28}))
\begin{eqnarray}
{\cal E}_{{\rm ex}}\equiv\sum\limits_l
\epsilon_l(\nu_{l+}+\nu_{l-})=\Delta_0
\sum\limits_l\left\{1+\frac{{\rm cosh}(\epsilon_l/2T)}
{{\rm cosh}(\widetilde\mu/2T)}\right\}^{-1}
\label{39a}
\end{eqnarray}
at $\bar\mu=\varepsilon$.
The entropy $S\equiv-\partial\Omega/\partial T$ is reduces to
the configuration form type of Eq. (\ref{26a}):
\begin{equation}
S=-\sum\limits_l\left\{\left[\nu_{l+}\ln\nu_{l+}+
(1-\nu_{l+})\ln(1-\nu_{l+})\right]+
\left[\nu_{l-}\ln\nu_{l-}+
(1-\nu_{l-})\ln(1-\nu_{l-})\right]
\right\}.
\label{26b}
\end{equation}

In the simplest case $\bar\mu=\varepsilon$ equations (\ref{37}),
(\ref{39}) govern the long--range ordering to be reduced to the form
\begin{eqnarray}
\tanh {\eta+\widetilde m\over\Theta}+
\tanh {\eta-\widetilde m\over\Theta}&=&2\eta,
\label{37c}\\
\tanh {\eta+\widetilde m\over 2\Theta}-
\tanh {\eta-\widetilde m\over 2\Theta}
&=&2(2\overline C - 1)
\label{39c}
\end{eqnarray}
\end{widetext}
where dimensionless values of the
sublattice difference of the chemical potentials
$\widetilde m\equiv\widetilde\mu/\Delta_0$ and temperature $\Theta\equiv
T/T_c$ are introduced. Analytical consideration of this system is
possible in the limits $\Theta\ll 1$ and $\eta\ll 1$ only. In the
first case, with accuracy of the first non--vanished terms,
equations (\ref{37c}), (\ref{39c}) arrive at the expressions
\begin{eqnarray}
&\eta=\eta_0-\exp\left(-{\Delta_0\eta_0\over T}\right),&\\
&\widetilde\mu =\Delta_0\eta_0+\alpha T&
\label{41}
\end{eqnarray}
where parameters $\eta_0$, $\alpha$ are determined by relations
\begin{eqnarray}
\eta_0={1\over 2}\left(1-\tanh {\alpha\over 2}\right)=
\frac{(1-\overline C)^2}{{1\over 8}+2(\overline C-{3\over 4})^2}.
\label{41a}
\end{eqnarray}
The phase transition curve where
\begin{equation}
\eta=0,\ {\rm cosh}{\widetilde m\over\Theta}=\frac{1}{\sqrt{\Theta}}=
\frac{1+\delta^2}{1-\delta^2};\quad\delta\equiv 2\overline C-1
\label{42a}
\end{equation}
is determined by relations
\begin{equation}
2\overline C-1 = {\sqrt{T_c-T}\over \sqrt{T_c}+\sqrt{T}},\quad
T=T_c\left(\frac{1-\delta^2}{1+\delta^2}\right)^2.
\label{42b}
\end{equation}
In the vicinity of the transition curve $T_c(\delta)$ one has
\begin{eqnarray}
&\eta^2=A(\delta)t(\delta),\quad t(\delta)\equiv
{\frac{T_c(\delta)-T}{T_c(\delta)}};&\label{42cc}\\
&{\rm cosh}{\widetilde\mu\over 2T}=\frac{1+\delta^2}{1-\delta^2}+
\frac{A(\delta)\delta^2}{1-\delta^2}t(\delta)&
\label{42c}
\end{eqnarray}
where factor $A(\delta)$ is determined by the equality
\begin{eqnarray}
A^{-1}\equiv\frac{2\delta^2}{1+\delta^2}+
\left[\left(\frac{1-\delta^2}{1+\delta^2}\right)^2-{2\over 3}\right]
\left(\frac{1+\delta^2}{1-\delta^2}\right)^4.
\label{42d}
\end{eqnarray}
With deviation of the stoichiometric composition this factor decays as
$A\simeq\sqrt{3}(1-2\delta^2)$, $\delta^2\ll 1$ taking zeroth
magnitude at $\delta_0=0.364$ ($C_0=0.682$). Physically,
this means that with passing out of the domain $0.318<\overline C<0.682$
the second order phase transition is transformed into the first
one being close to the second.

\begin{figure}[htb]
\caption{Temperature dependencies of the long--range order parameter (a) and
the sublattice difference of the chemical potentials (b)
at different concentrations $\overline C$ being pointed out
near corresponding curves. (The dashed line bounds the abruption
region)}
\end{figure}
Explicit forms of the dependencies $\eta(T,\overline C)$,
$\widetilde\mu(T,\overline C)$ in the case $\bar\mu-\varepsilon=0$
are depicted in Figures 3, 4.
The order parameter $\eta$ decays monotonically with
increase of both temperature and concentration (see Figures 3a, 4a),
whereas behavior of the sublattice difference of the chemical
potentials $\widetilde\mu$ is much more complicated: when the
temperature grows, it decreases near the stoichiometric composition
$\overline C=0.5$ passing to increasing regime with approaching to
the composition $\overline C=1$ (Figure 3b); analogous behavior is
observed with deviation of the stoichiometric composition at low and
pre--critical temperatures (Figure 4b). Principle important is the
interruption of the sublattice difference of the chemical
potentials $\widetilde\mu$ at the curve of the phase transition
that takes maximal values within intermediate ordering region.
\begin{figure}[htb]
\caption{Compositional dependencies of the long--range order parameter (a) and
the sublattice difference of the chemical potentials (b)
at different temperatures (the magnitudes of the latter related
to critical value are pointed out near corresponding curves).
The dashed line bounds the abruption region}
\end{figure}

If the system is subjected to external influence, the parameter
$\bar m\equiv(\bar\mu-\varepsilon)/\Delta_0$ becomes non--zeroth
and ordering process is described by the system (\ref{37c}),
(\ref{39c}) where the order parameter $\eta$ should be replaced by
the renormalized one $\sqrt{\eta^2+\bar m^2}$. It is easy to
convince oneself the growth of the influence parameter $\bar m$
causes suppressing ordering process to shrink the long--range
domain as shown in Figure 5a. At that, the boundary composition
being the solubility limit of the ordering phase is determined by
the parameter of external influence in following manner:
\begin{equation}
\overline C = 1-\frac{\sqrt{\bar m(1-\bar m)}-\bar m}{2(1-2\bar m)},\quad
\bar m\equiv{\bar\mu-\varepsilon\over\Delta_0}.
\label{42f}
\end{equation}
According to related plot in Figure 5b the solubility concentration
decays monotonically with arising external influence to
vary anomalously fast near boundary magnitudes $\overline C = 1$ and
$\overline C = 0.5$.
\begin{figure}[htb]
\caption{a --- The phase diagram determining long--range order (LRO) and
disorder (DO) domains at different magnitudes of chemical potentials
(curves top-down correspond to values
$(\bar\mu-\varepsilon)/\Delta_0=0.0,\ 0.2,\ 0.3,\ 0.4,\ 0.5,\ 0.6,\ 0.7,\
0.8,\ 0.9$). b --- The solubility limit of the ordering phase versus
the parameter of external influence}
\end{figure}

\section{Collective excitations}\label{sec:level5}

As is seen from Sections \ref{sec:level2} and \ref{sec:level4}, the
ordering process results from the fact that, as the temperature
decreases, the effective Fermions form coupled pairs
corresponding to the collective excitations of the Bose type.
If the number of such pairs ${\cal N}={1\over 2}N\eta$ makes up the finite part
with respect to the total number of sites $N$, the ordering
process assumes the macroscopic character and is determined by the
long--range order parameter $\eta=\Delta/\Delta_0$.
So, in the representation of single excitations,
the solid solution ordering is reflected
through the appearance of the off--diagonal components
$\Delta$ in the matrix of the self--energy function
$\widehat\Sigma$.

Let us show now how the ordering manifests itself in the
representation of the collective excitations. Application of the
mean--field method, therewith, turns to be insufficient and
a recourse should be made to the self--consistent scheme developed
in Section \ref{sec:level3} to take into account the behavior of
the Green functions of both single and collective excitations. To
study the latter we start with consideration of the Fourier
representation $\widehat\Gamma ({\bf K},\Omega)$ of the vertex
function acquiring under ordering a condensate component
$\widehat{\Gamma}_0({\bf K},\Omega)$ accompanied by a fluctuation
function $\widehat{\Gamma}'({\bf K},\Omega)$. Their principle
difference is in hydrodynamic behavior in the limit ${\bf
K},\Omega\to 0$ where the latter tends to zero, whereas the former
takes a finite magnitude $\widehat{\Gamma}_0\ne 0$. For
determination of the latter we substitute the polarizer
(\ref{30}) into (\ref{12c}):
\begin{widetext}
\begin{eqnarray}
\widehat{\Gamma}_0^{-1}=
\left(\begin{array}{cc}
B-A|\Delta|^2 &A\Delta ^2\\
A(\Delta ^*)^2 &B-A|\Delta|^2
\end{array}\right),
\quad B\equiv -W^{-1}+
{1\over 2\epsilon}\tanh {\epsilon\over 2T}
\label{73}
\end{eqnarray}
\end{widetext}
where account is taken of the diagonal
structure of the interaction matrix $\widehat{w}$, the parameter $A$
is determined by the last equality (\ref{31})
(in this Section, the gap $\Delta$ is supposed to be complex).
It is principally important the parameter $B\to 0$ with ordering,
in accordance with condition Eq. (\ref{21}) for the gap $|\Delta|$.

Inversion of the matrix (\ref{73}) arrives at the off--diagonal
component of the condensate vertex:
\begin{eqnarray}
\Gamma_0^{01}=-\Delta_0\frac
{{1\over\eta}\tanh {\eta\over\Theta}-
{1\over\Theta}{\rm cosh}^{-2}{\eta\over\Theta}}
{{\left(1-{1\over\eta}\tanh {\eta\over\Theta}\right)}
{\left(1-{1\over\Theta}{\rm cosh}^{-2}{\eta\over\Theta}\right)}}.
\label{73a}
\end{eqnarray}
Near the critical point $(T-T_c\ll T_c)$ where
\begin{eqnarray}
\Gamma_0^{01}\simeq -{2\over 3}\Delta_0\frac{\eta^2}{(1-T/T_c)^2},
\label{73b}
\end{eqnarray}
we can neglect a frequency dispersion to put
$\Sigma^{01}(\Omega)\simeq\Sigma^{01}(0)$,
$\Gamma^{01}(\Omega)\simeq\Gamma^{01}(0)\equiv\Gamma^{01}_0$ in
Fourier transform of the equality (\ref{12b})
that takes the form
\begin{eqnarray}
{1\over\eta}\tanh {\eta\over\Theta}=-2\frac{\Delta_0}{\Gamma_0^{01}}.
\label{73c}
\end{eqnarray}
As a result, the collective excitations are characterized by
the relations
\begin{eqnarray}
\eta\simeq\sqrt{3}\left(1-\frac{T}{T_c}\right),\
\Gamma_0^{01}\simeq -2\Delta_0;\quad T-T_c\ll T_c
\label{73d}
\end{eqnarray}
the first of which is principally different of the corner peculiarity
(\ref{42cc}) inherent in the single excitations.
This difference is expressed physically in that the collective excitations
polarize ordering system to transform the second order phase transition
into the first one.

Now, we are in position to study the collective excitations themselves.
Their Green function is determined by the fluctuation component
$\widehat{\Gamma}'({\bf K},\Omega )$ of the vertex function
according to the relation
\begin{eqnarray}
\widehat{\varphi}'({\bf K},\Omega)\equiv
-\frac{1}{2\pi{\rm i}}
\widehat{\Gamma}'({\bf K},\Omega).
\label{73e}
\end{eqnarray}
The simplest way to find this function in hydrodynamic limit
${\bf K},\Omega \to 0$ is to
apply the method developed in Refs. \cite{20}, \cite{22}.
Its main point consists in the fact that instead of the equation
$\widehat{G}^{-1}=\widehat{G}^{-1}_0-\Delta_0\widehat{\delta}$,
for the Green function $\widehat{G}$, use is made of
its analog $\left< \widehat{G}\right>^{-1}=\widehat{\sigma }^{-1}-
\Delta _0\widehat{\delta}$ \ for the averaged function $\left<
\widehat{G}\right>$ and locator
$\widehat{\sigma }=\left< \widehat{G}_0\right>$.  In addition,
inserted is the effective interactor $\widehat{U}= \Delta
_0\widehat{\delta }+\Delta _0^2 \left<\widehat{G}\right>$ which
contains the term in the first order over Fermion energy.
The Dyson equation replacing
(\ref{12a}) therewith assumes the form  $\widehat{U}^{-1}= \Delta
_0^{-1}\widehat{\delta }-\widehat{\sigma }$.
The two--particle Green function
\begin{widetext}
\begin{equation} \varphi (E;{\bf K},\Omega )
=-{1\over 2\pi{\rm i}} \sum_{{\bf k},{\bf k}'} \left< G^{\rm
R}({\bf k}_+,{\bf k}'_+; E+\Omega )G^{\rm A}({\bf k}_-,{\bf k}'_-;
E)\right> \label{74} \end{equation} appears, in the ladder
approximation, as (compare with (\ref{12c}))
\begin{equation} \varphi
(E;{\bf K},\Omega )=-{1\over 2\pi{\rm i}} \left[ \gamma ^{-1}-\sum _{\bf
k}U^{\rm R}({\bf k}_+; E+\Omega )U^{\rm A}({\bf k}_-; E)\right]
^{-1} \label{75}
\end{equation}
where ${\bf k}_\pm={\bf k}\pm {\bf
K}/2$, $\gamma$ is an irreducible four--tail vertex, indices R
and A of the retarded and advanced functions correspond to the
selection of the components $\alpha =0$,~1 in various sublattices.
Hence, taking into account the Dyson equation and
Ward identity \cite{9}, we have
\begin{equation} \sigma _{{\bf
k}_+}^{\rm R}(E+\Omega )-\sigma _{{\bf k}_-}^{\rm A}(E)= \gamma \sum
_{{\bf k}'} \left[ U^{\rm R}({\bf k}'_+; E+\Omega )-U^{\rm A}({\bf
k}'_-; E) \right] -\widetilde\gamma \, \Omega\> \label{76}
\end{equation}
\end{widetext}
where the irreducible vertex $\widetilde\gamma$ has,
contrary to the $\gamma$, two coinciding tails appropriate to the
same sites. Then, we come to the conventional expression for the
fluctuation component of the two--particle Green function
\begin{equation} \varphi '({\bf
K},\Omega )=-{\chi ({\bf K})\over \Omega +{\rm i}D({\bf K},\Omega ){\bf
K}^2}
\label{77}
\end{equation}
where $\chi ({\bf K})$ is susceptibility taking
in hydrodynamic limit ${\bf K}\to 0$ the thermodynamic value $\chi$,
$D({\bf K},\Omega )$ is the dispersing diffusion coefficient.
As a result of the above, for the fluctuation component of the vertex
function we obtain \cite{20,22}
\begin{eqnarray}
\left(\widehat{\Gamma}'({\bf K},\Omega )\right)^{-1}
={1\over 2\pi \chi}
\left(\begin{array}{cc}
-{\rm i}\Omega +D{\bf K}^2 &0\\
0 &{\rm i}\Omega+D{\bf K}^2
\end{array}\right)
\label{79}
\end{eqnarray}
where the matrix structure reflects the presence
of two poles $\Omega= \mp {\rm i}D{\bf K}^2$.

The sum of the expression (\ref{73}) with $B = 0$ and Eq. (\ref{79})
produces, after the matrix inversion, the complete Green function of
collective excitations:
\begin{widetext}
\begin{eqnarray}
&\widehat{\varphi} ({\bf
K},\Omega )={\chi\over {\cal D}({\bf K},\Omega )}
\left(\begin{array}{cc}
-\Omega -{\rm i}\left({1\over 2}SK_0-D{\bf K}^2\right)
&-{{\rm i}\over 2}(\Delta /|\Delta |)^2SK_0\\
-{{\rm i}\over 2}(\Delta ^*/|\Delta |)^2SK_0
&\Omega -{\rm i}\left({1\over 2}SK_0-D{\bf K}^2\right)
\end{array}\right),&
\label{80a}\\
&{\cal D}({\bf K},\Omega )=\Omega ^2+\left( (1/2)SK_0-D{\bf
K}^2\right) ^2 -(1/4)S^2K_0^2,&\label{80b}\\
&S^2=4\pi \chi A|\Delta
|^2D,\qquad K_0^2=4\pi \chi A|\Delta |^2/D.&\label{80d}
\end{eqnarray}
\end{widetext}
\noindent Condition ${\cal D}({\bf K},\Omega )=0$ brings to the
dispersion law
\begin{equation}
\Omega =\mp DK\sqrt{K_0^2-{\bf K}^2}\>
\label{81}
\end{equation}
whose characteristic feature consists in a bell form with a slanting
long--wave side (see Figure 6). Physically,
the dependence (\ref{81}) relates to the collective mode that is of a reactive nature
in the long--wave region $K<K_0$ limited by
the wave number given Eq. (\ref{80d}) and has a relaxation nature in the
short--wave region $K>K_0$. At $K\gg K_0$, it transforms
into the ordinary diffusion mode $\Omega =\mp {\rm i}D{\bf K}^2$.
\begin{figure}[htb]
\caption{Law of collective mode dispersion (the solid line
corresponds to the real value of frequency $\Omega$, the dashed
--- imaginary value, the dotted --- diffusion mode)}
\end{figure}

In this way, the self--consistent consideration of single and
collective excitations leads to the conclusion that apart from the
diffusion regime realized in the mesoscopic region $K>K_0$,
possible in the system is quasi--oscillations characterized by the
phase velocity $S$ determined by the first of equalities
(\ref{80d}). Obviously, the dispersion law (\ref{81}) acquiring
the acoustic form in the limiting long--wave region $K\ll K_0$ is
observed as Zener peak in experiments with internal friction
\cite{Postnikov}. This mode is characterized by the frequency
\begin{equation}
\Omega_0\equiv SK_0={\pi\chi\over\Delta_0}
\left({1\over\eta}\tanh {\eta\over\Theta}-
{1\over\Theta}{\rm cosh}^{-2}{\eta\over\Theta}\right)
\label{81a}
\end{equation}
whose temperature dependence
at concentration $\overline C=0.5$ is shown in Figure 7 with solid line.
This frequency is seen to decay monotonically from
the magnitude $\Omega_0=\pi\chi/\Delta_0$ at $T=0$ to $\Omega_0=0$ at $T=T_c$
varying near the critical point quadratically:
\begin{equation}
\Omega_0\simeq 2\pi{\chi\over\Delta_0}
\left(1-\frac{T}{T_c}\right)^2,\quad T-T_c\ll T_c.
\label{81b}
\end{equation}
Here, we take into account polarization effects transforming
the corner peculiarity (\ref{42cc}) into the linear relation (\ref{73d}).
\begin{figure}[htb]
\caption{Temperature dependencies of the characteristic frequency
$\Omega_0$ (solid line), phase velocity $S$ (broken
line) and boundary value of the wave number $K_0$ (dotted line).
Magnitudes $\Omega_0$, $S$ and $K_0$ are measured in units $\pi\chi/\Delta_0$,
$\sqrt{\pi\chi D_0/\Delta_0}$ and $\sqrt{\pi\chi/\Delta_0 D_0}$, respectively.
The height of diffusion barrier is put to be $Q=3T_c$.
The insertion is built with accounting polarization effects given by
Eqs. (\ref{81b}) --- (\ref{81cc})}
\end{figure}
Since the diffusion coefficient $D=D_0\exp\{-(Q/T)\}$ is determined
by the energy barrier $Q$ to behave itself
in a non--critical manner, the parabolic dependence (\ref{81b})
means the linear behavior
for both the sound velocity and the boundary wave number:
\begin{eqnarray}
&&S\simeq\sqrt{2\pi{\chi D_0\over\Delta_0}}\exp\left(-{Q\over 2T_c}\right)
\left(1-\frac{T}{T_c}\right),\label{81c}\\
&&K_0\simeq\sqrt{2\pi{\chi\over\Delta_0 D_0}}\exp\left({Q\over 2T_c}\right)
\left(1-\frac{T}{T_c}\right)
\label{81cc}
\end{eqnarray}
the former of which is exponentially smaller than the latter.
With the temperature falling down in the domain $T\ll T_c$
the sound velocity decreases and the boundary value of the wave number,
on the contrary, infinitely increases:
\begin{eqnarray}
&&S\simeq\sqrt{\pi\chi D_0\over\Delta_0}\exp\left(-{Q\over 2T}\right),
\label{81d}\\
&&K_0\simeq\sqrt{\pi\chi\over\Delta_0 D_0}\exp\left({Q\over 2T}\right).
\label{81dd}
\end{eqnarray}
A concentration deviation off the
stoichiometric composition causes order suppressing to decrease the
values $\Omega_0$, $K_0$ and $S$. It follows therefrom that the most
preferable (in terms of detection of the zero--sound mode of collective
excitations) is the temperature region that is situated just below the
critical temperature $T_c$. Explicit form of the temperature dependencies
$\Omega_0(T)$, $S(T)$ and $K_0(T)$ are depicted in Figure 7
where main panel relates to the phase transition of the second order,
whereas insertion takes into account polarization effects
transforming it into the first order.

According to Eqs. (\ref{80a}) --- (\ref{80d}) the diagonal components
of complete vertex function have the following hydrodynamic form:
\begin{eqnarray}
\Gamma^{\alpha\alpha}({\bf K},\Omega_S)=
2\pi\chi\frac{\Omega_S+{1\over 2}\Omega_0}
{\Omega_S^2+\Omega_0 D{\bf K}^2},
\label{84}\\
\Omega_S\equiv 2\pi ST,\quad
S=0,\pm 1,\ldots
\nonumber
\end{eqnarray}
Then, taking into account the sum rule \cite{43} (cf. Eqs. (\ref{36}))
\begin{equation}
T\sum_{S=-\infty}^{\infty}\frac{1}{\Omega_S^2+\epsilon^2}=
{1\over 2\epsilon}\coth {\epsilon\over 2T},
\label{85}
\end{equation}
for instant vertex function $\Gamma^{\alpha\alpha}({\bf K})\equiv
T\sum_S\Gamma^{\alpha\alpha}({\bf K},\Omega_S)$ we obtain
\begin{equation}
\Gamma^{\alpha\alpha}({\bf K})={\pi\over 2}\chi\sqrt{\Omega_0\over D{\bf K}^2}
\coth {\sqrt{\Omega_0 D{\bf K}^2}\over 2T}.
\label{86}
\end{equation}
In the hydrodynamic regime $K\ll T/\sqrt{\Omega_0 D}$ this expression
arrives at the known Bogolyubov singularity \cite{35}
\begin{equation}
\Gamma^{\alpha\alpha}({\bf K})=\pi{\chi T\over D{\bf K}^2}
\label{87}
\end{equation}
that keeps this form for off--diagonal components of the vertex function
as well.

\section{Conclusions}\label{sec:level6}

Above consideration shows that within the single excitation
representation the study of the solid solution ordering can be
achieved by analogy with the microscopic theory of
superconductivity \cite{29}, \cite{24}. Description of the
collective excitations requires their self--consistent
consideration along with single ones. The hydrodynamic behavior of
the system is presented as the result of interference of
the condensate and fluctuation components the last of which is of
diffusive type. This interference results in appearance of the
reactive mode corresponding to the zero--sound.

It is naturally, the scheme developed allows one to reproduce
complete picture of the solid solution ordering within mean--field
approximation \cite{a} --- \cite{D}. Setting aside well--known information
along this approach, let us focus on main results of our consideration that
are not achieved within  mean--field scheme.

First, we have found the temperature dependence of the long--range order
parameter for different differences of the chemical potentials of the components
$\mu\equiv\mu_A-\mu_B$ accounted from Fermion energy $\varepsilon$ (see Figure 2).
Taking the latter as the origin, we obtain a scope of ordering solid solutions
is determined by the condition
\begin{equation}
\mu_A-\mu_B<\Delta_0\equiv W/2
\label{30Aa}
\end{equation}
fixing degree of the chemical affinity of the components with respect to
the characteristic value $W\equiv -\sum_m w_{lm}$
of mixing energy (\ref{6}).
In accordance with Eq. (\ref{30a}), it means the difference of sublattice
concentrations $2\widetilde C\equiv C_1-C_0$ could not take magnitudes
more than one.

The second of practical results allows us to determine the boundary composition
of the ordering phase $AB_n$:
\begin{equation}
\overline C =\frac{\sqrt{\bar m(1-\bar m)}-\bar m}{(n+1)(1-2\bar m)},\quad
\bar m\equiv{\mu_A-\mu_B\over\Delta_0}.
\label{42g}
\end{equation}
As distinct from equality (\ref{42f}) related to the simplest phase $AB$,
here we consider the concentration domain $0\leq\overline C\leq(n+1)^{-1}$
instead of $0.5\leq\overline C\leq 1$.
According to relation (\ref{42g}), in perfectly ordering solution
where $\mu_A=\mu_B$, the solubility limit is $\overline C=0$ and
grows anomalously fast with decreasing degree of the chemical affinity of the
components with respect to the mixing energy.
Moreover, if we insert impurities or defects with
concentration $x\ll 1$, the parameter of external influence
varies as
\begin{equation}
\bar m=m_0+\beta x
\label{42h}
\end{equation}
where the value $m_0$ is inherent in the proper solution $AB_n$,
$\beta$ is constant taking both positive and negative
signs. Thus, alloy doping arrives at lowering boundary
concentration in the case $\beta<0$ and its growth otherwise.

Principle advantage of our approach is a possibility to study
solid solutions with arbitrary composition
to elaborate complete thermodynamic picture of
the long--range ordering process.
According to Figures 3, 4 the order parameter decays monotonically with
increase of both temperature and concentration, whereas the sublattice
difference of the chemical potentials suffers abruption with
maximal value $\widetilde\mu\sim\Delta_0/2$ at phase transition related to
intermediate compositions. This allows us to estimate the surface tension
coefficient $\sigma\sim\widetilde\mu\xi/v$ as follows:
\begin{equation}
\sigma=c{W\xi\over v},\quad c\sim 0.1
\label{90}
\end{equation}
where $\xi$, $v$ are characteristic values of the correlation
length and the atom volume, respectively. Another physical result is that the
second order phase transition is transformed into the first one
with passing out of the compositional domain $0.318<\overline
C<0.682$. However, a difference between these transitions is so
weak to be observed in Figure 3a.

Much more important is the influence of collective excitations
whose polarization effects transform the phase transition to the
first order always (see the first of Eqs. (\ref{73d})).
Related collective mode is of a reactive nature in the long--wave
region limited by the wave number (\ref{80d}).
The dispersion law (\ref{81}) of this mode can be observed as
the Zener peak of the internal friction.
According to Figure 7 proper frequency and boundary wave number
of this peak decay monotonically with temperature increase, whereas
characteristic velocity has a maximum at intermediate temperatures in ordering domain.
A deviation off the stoichiometric composition causes suppressing
of the above zero--sound mode to be pronounced in the temperature region
just below the critical temperature.
It is worthwhile to notice
the polarization effects are relevant to the static (condensate)
component of Green function (\ref{80a}) related to frequency $\Omega=0$,
whereas the Goldstone mode of the symmetry restoration is represented by
the instant vertex function (\ref{87}) related to time $t=0$.

\section*{Acknowledgments}\label{sec:level1}

In this work, financial support by STCU, project 1976,
is gratefully acknowledged.


\begin{thebibliography}{00}

\bibitem{a} A.D. Bruce, R.A. Cowley, {\it Structural Phase Transitions}
(Taylor and Francis Ltd., London, 1981).

\bibitem{b} M.A. Krivoglaz, {\it The Theory of X--Ray and Thermal Neutron
Scattering From Real Crystals} (Plenum, New York, 1969);
{\it Diffuse Scattering of X--rays and Thermal Neutrons by Fluctuational
Inhomogeneities of Imperfect Crystals} (Springer, Berlin, 1996).

\bibitem{c} D. de Fontaine, in: {\it Phenomena in Alloys,
Magnets, and Superconductors}, ed. R.E. Millis (McGraw-Hill, New York, 1971);
{\it Solid State Physics} {\bf 34}, 73 (1979).

\bibitem{d} A.G. Khachaturyan, {\it Theory of Structural Transformations
in Solids} (Wiley, New York, 1983).

\bibitem{KO} A.A. Katsnelson, A.I. Olemskoi, {\it Microscopic Theory of
Nonhomogeneous  Structures}, (Mir Publishers, Moscow, 1990).

\bibitem{D} F. Ducastelle,  {\it Order and Phase Stability in Alloys}
(Elsevier, New York, 1991).


\bibitem{9} J. Zinn-Justin, {\it Quantum Field Theory and
Critical Phenomena} (Clarendon Press, Oxford, 1993).


\bibitem{35} N.N. Bogolyubov, {\it Quasi--mean Values in the Problems
of Statistical Mechanics. In Coll. Statist. Phys. and Quantum Theory
of Field} (Nauka, Moscow, 1973).

\bibitem{37} L.D. Landau, E.M. Lifshits, {\it Course of Theoretical Physics}
(Pergamon, Oxford, 1980), Vol. 5.

\bibitem{43} A.P. Prudnikov, Yu.A. Brychkov, O.I. Marichev, {\it
Integrals and Series} (Nauka, Moscow, 1981).

\bibitem{20} D. Vollhardt, P. W\"olfle, {\it Phys. Rev.} B{\bf 22}, 4666
(1980).

\bibitem{22} T. Kopp, {\it J. Phys.} C{\bf 17}, 1897, 1919 (1984).

\bibitem{Postnikov} V.S. Postnikov, {\it Internal Friction in Metals}
(Metallurgiya, Moscow, 1974).

\bibitem{29} G. Shriffer, {\it Theory of Superconductivity} (Nauka,
Moscow, 1979).

\bibitem{24} R. Haussmann, {\it Z. Phys.} B{\bf 91}, 291 (1993).

\end{thebibliography}
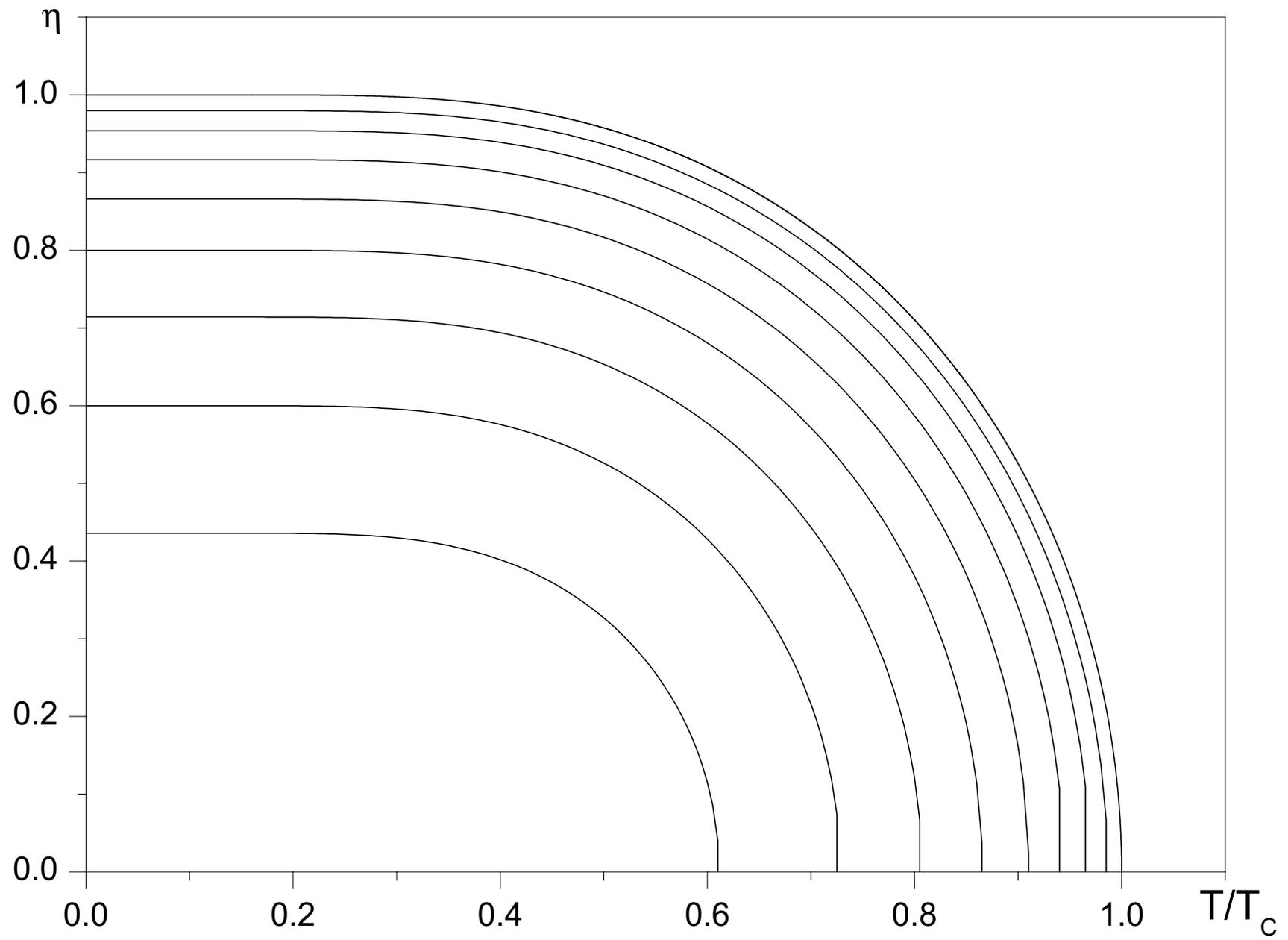
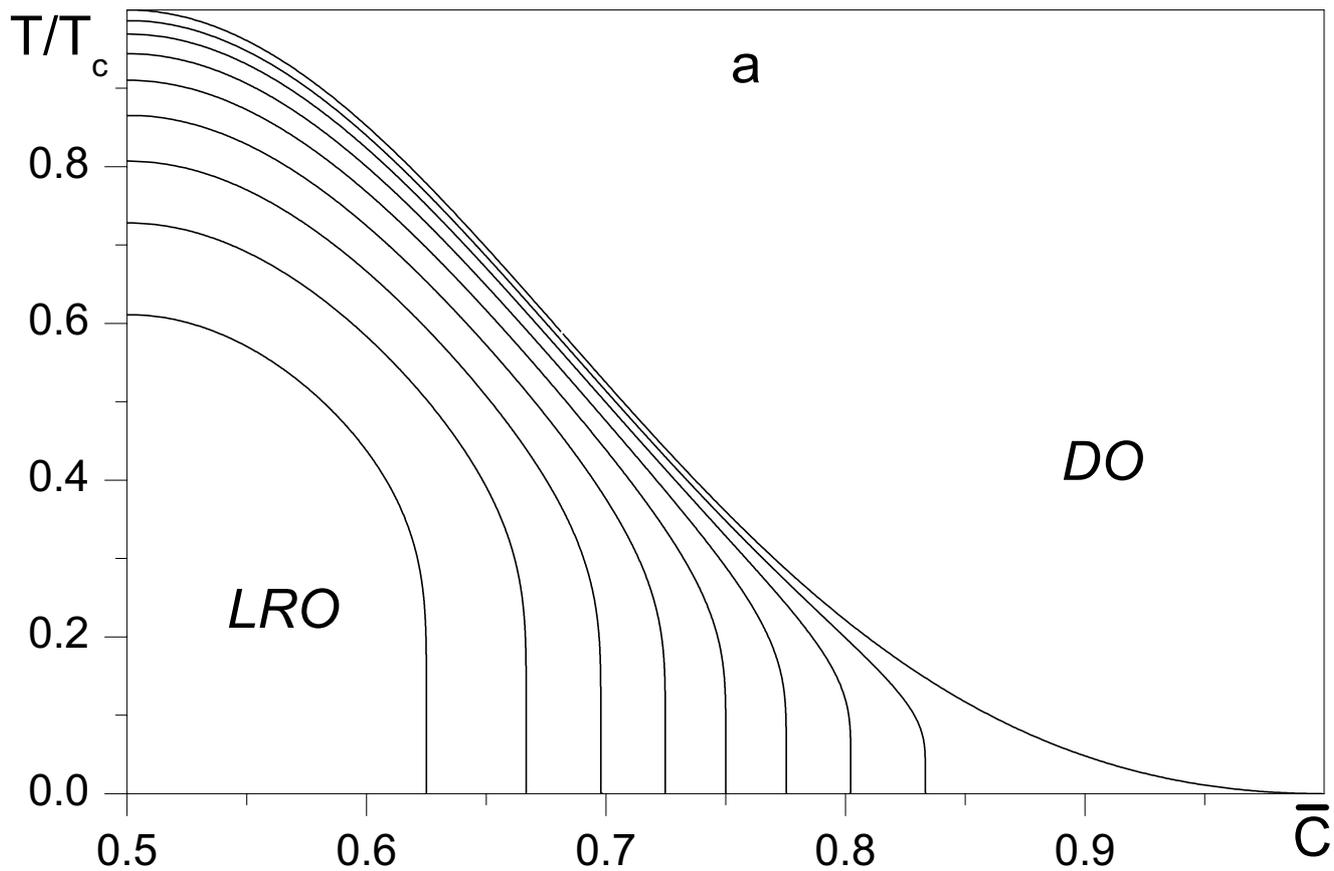
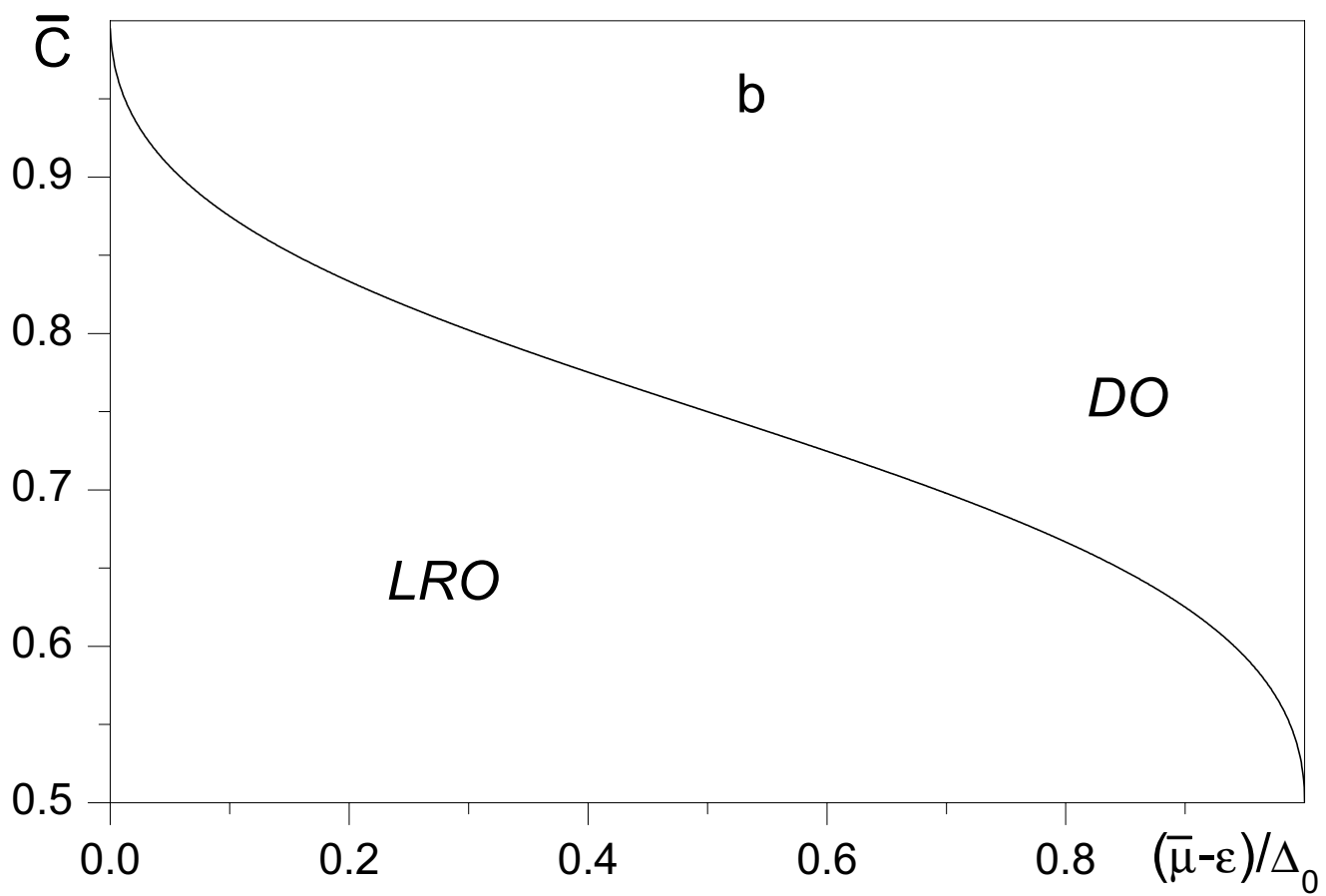
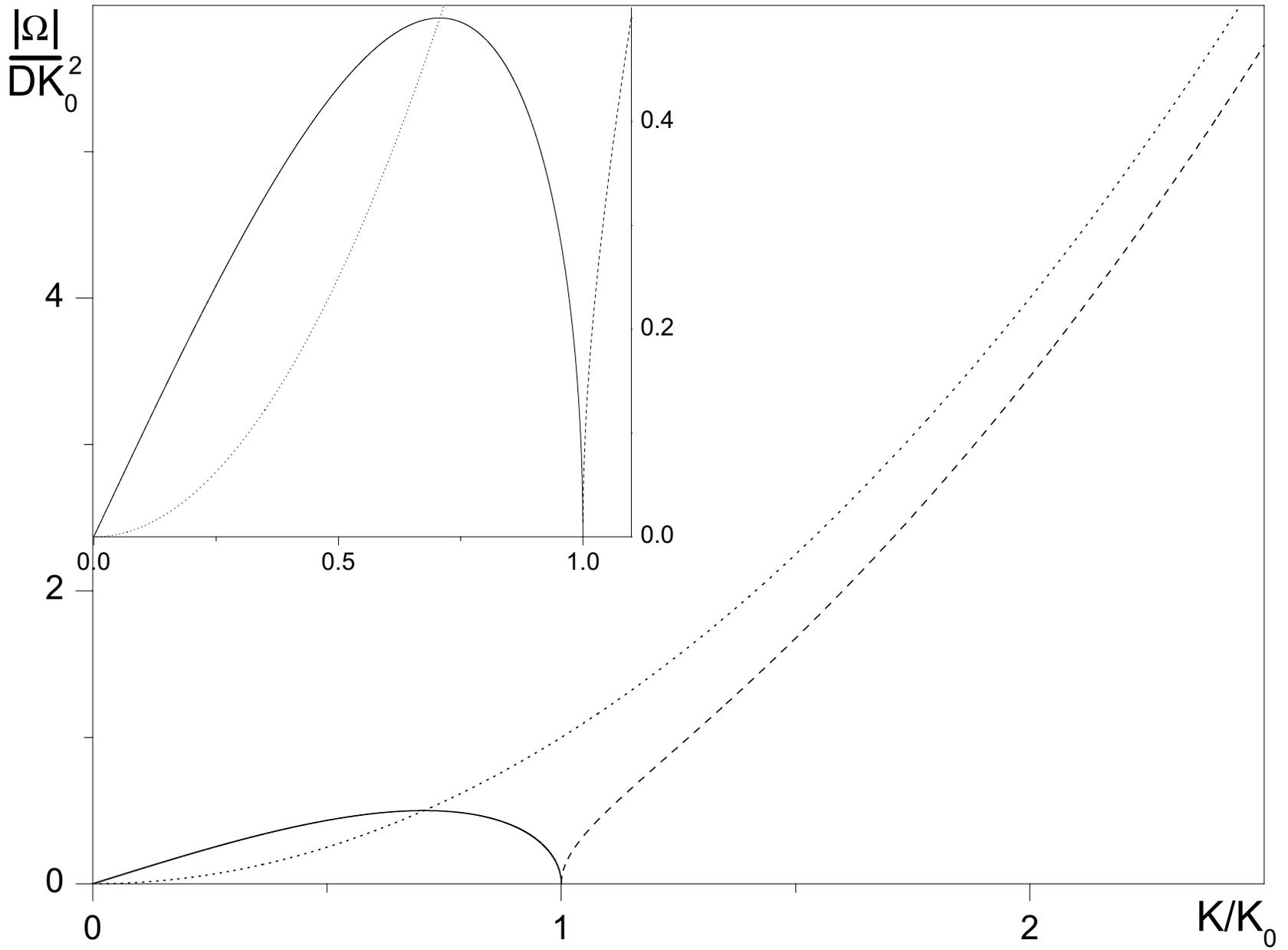
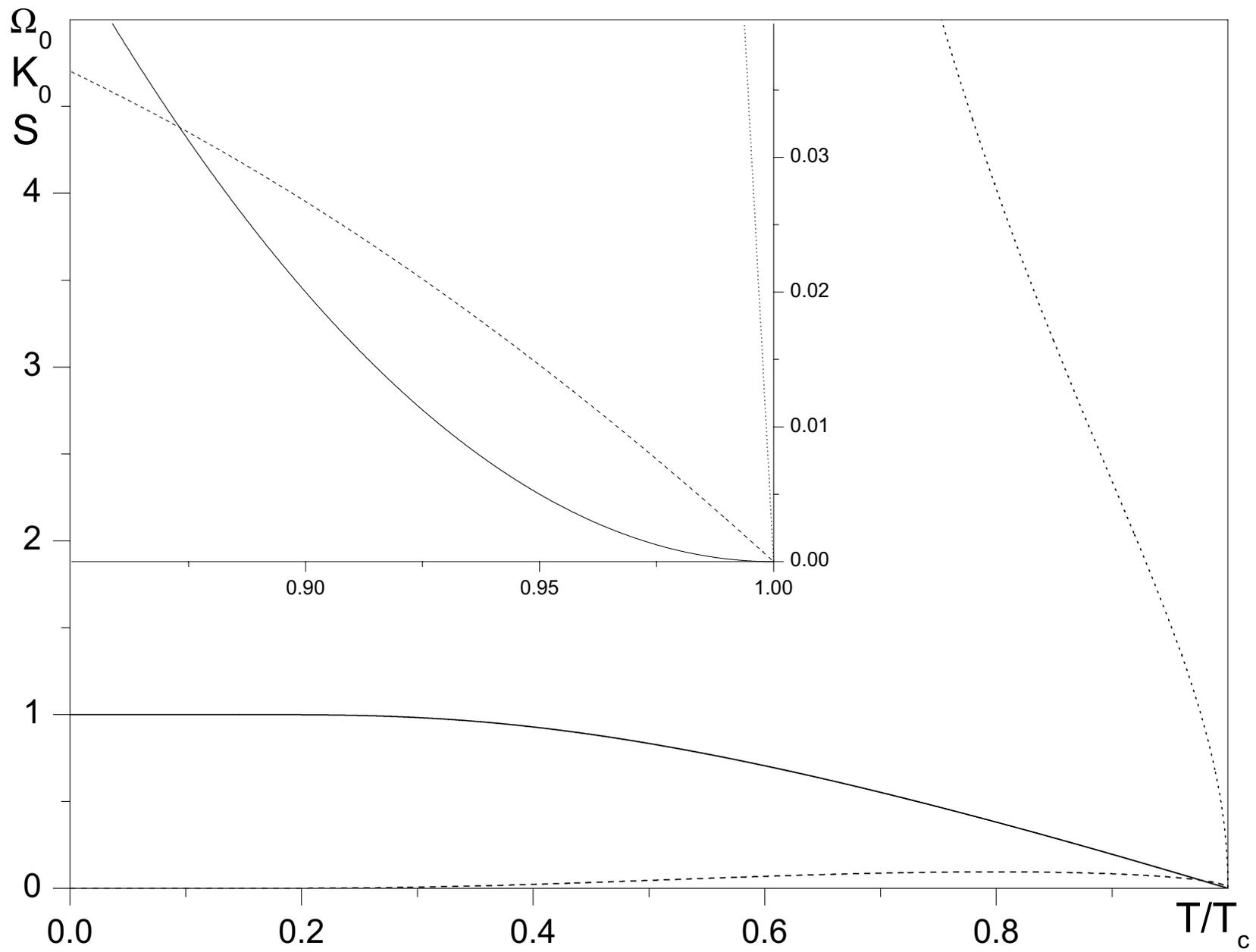
\end{document}